# Optimization of the metal-dielectric metasurface unit cell for sensitivity enhancement in determination of IgG concentration in solutions

K. S. Kuznetsova[1], V. A. Pashynska[1,2], Z. E. Eremenko[1,3]

[1] O.Ya. Usikov Institute for Radiophysics and Electronics of the NAS of Ukraine, 12 Academ. Proskury Str., Kharkiv 61085, Ukraine, tkachenkok89@gmail.com, zoya.eremenko@gmail.com

[2] B.Verkin Institute for Low Temperature Physics and Engineering of NAS of Ukraine, 47 Nauky Ave., Kharkiv, 61103, Ukraine, vlada.pashynska@gmail.com

[3] Leibniz Institute for Solid State and Materials Research, 01069 Dresden, Germany,

**Abstract.** This study focuses on developing a metal-dielectric sensor structure with optimized unit cell geometry for determination of protein Immunoglobulin G (IgG) concentration in aqueous solutions. The research combines both experimental and theoretical investigations, utilizing the differential microwave dielectrometry method and numerical modeling with COMSOL software. Complex permittivity (CP) values dependence of IgG water solutions on the protein concentration was experimentally obtained at the microwaves using original microwave dielectrometer setup. It was shown that increase of IgG concentration resulted in the CP values of the solutions studied decrease. The experimentally obtained CP data for the IgG water solutions were used as a basis for microwave metal-dielectric metasurface unit cell numerical modeling. The metal-dielectric metasurface consisting of Teflon substrate and plane copper microresonators was combined with a standard 96-well microplate used in clinical laboratories. Optimization of the obtained metasurface unit cell revealed that the size and position of the copper microresonators within the unit cell significantly impact the sensor sensitivity for determining the IgG concentration in aqueous solutions. The metasurface with the unit cell containing four copper microresonators provided the most sensitive platform for detecting variations in the IgG concentration in the sample. The frequency shift of the reflection coefficient was directly related to changes in the protein concentration. The calibration graph was developed for effective determination of IgG concentrations in the aqueous solutions.

# INTRODUCTION

Immunoglobulins G (IgG) is the major class of antibodies found in human blood and extracellular fluids [1], which plays a significant role in humoral immunity response to infections and in regulations of allergic reactions in a human organism. IgG concentration in serum and other biological liquid samples are an important indicator of the immune status of a patient. Such a determination of the IgG level is commonly used in biomedical diagnostic of different infectious diseases and other pathologies. Traditionally, enzyme-linked immunosorbent assay (ELISA) tests are applied for the IgG determination and quantification in the biological samples (blood, serum, etc.) in medical practice for the diagnostic purposes [2], [3]. The standard multi-well microplates are frequently used for detecting biological samples, making them a popular choice for immunoassays [4]. In general, ELISA is an analytical biochemical assay which is characterized by the high sensitivity and accuracy, but the specially developed and expensive enzyme immunoassay sets (with synthesized specific antibodies) are needed for this analysis. The development of the specific ELISA tests is time and cost consuming process, moreover diagnostic procedures by usage such ELISA tests require the biochemical laboratory conditions. That is why the new effective physical methods (which do not require of the antibodies and other biochemical reagents and equipment application) of the IgG level evaluation in the solutions are also in demand. Such physical methods and portative sensors developed on the basis of the physical methods can be considered as a cheaper and more robust alternative for the fast estimation of the IgG level in the liquid samples, including the biological liquids or technological solutions of IgG during pharmaceutical production of the immunoglobulin preparations.

Among the methods of identification and quantification of biological substances (including proteins such as immunoglobulins) in the solutions electromagnetic sensing techniques are characterized as very effective methods [5], [6]. In the microwave dielectric sensing techniques, the complex permittivity (CP) of the tested solution can be used as a parameter for determination of the amount of the biomolecules in the aqueous solutions [7-9]. Taking into account the fact that water molecules bound to the hydrated biomolecules in the solution have limited mobility and reduced polarizability it is clear why the increase of, for example, protein concentration in the tested solution results in the decrease in both the real and imaginary CP parts of the water solution of the protein. In the microwave range, the changes in the ratio of the free and bound water molecules in the solution depending on some proteins dissolved concentration was observed using microwave dielectric spectroscopy [10-14]. At the same time the new type of sensors for the biological substances determination in the solutions based on the usage of metamaterials [15], [16] and metasurfaces [7] attract the researcher's attention today because of the high selectivity, sensitivity and miniaturization of such sensors. Development of metasurfaces with high quality factor (Q-factor) is crucial for achieving strong field concentrations in subwavelength volumes. The demand for resonant metasurfaces with high Q-factor is particularly significant for applications involving the sensing of biomolecules in the samples [17].

The current research is aimed at the experimental determination of the dielectric permittivity of the IgG aqueous solutions by differential microwave dielectrometry as well as numerical modeling of the sensors based on metasurface with the purpose to develop the new methodical approach for IgG concentration determination in solutions and to optimize the metasurface unit cell structure for the enhancing the sensitivity of the metasurface based sensors. In this study, we explored the potential integration of the metasurface with a standard multi-well microplate which is frequently used in biochemical laboratory testing practice. This integration facilitates further miniaturization of the sensors for analytes concentration determination, provides reducing the amount of the testing liquid samples and offers novel approach for characterizing solutions within the microplate wells.

## MATERIALS AND METHODS

**Materials.** For the microwave dielectrometry measurements, diluted IgG solutions were prepared using a 5% IgG aqueous solution (50 mg/ml) provided by "Biopharma" company (Ukraine).

**Experiment.** The real and imaginary parts of the CP of the IgG in water solutions were determined experimentally by using the microwave differential dielectrometer at the fixed frequency of 31.82 GHz. This method is based on solving of the electrodynamic problem using Maxwell's equations for circular layered waveguide structures, as described in detail in [18, 19]. We selected the microwave frequency range for obtaining the dielectric properties of the IgG in water solutions because this range includes the region of maximum frequency dispersion of the real and imaginary CP parts of water. By using the dielectrometry method in the microwave range, it is possible to describe the state of water molecules (whether water molecules are free or bound to the analyte molecules) in aqueous solutions depending on the analyte concentration, that is related to such molecular effect in this frequency range as dipole relaxation of free water molecules.

**Modeling.** We performed the numerical modeling of a structure containing the standard 96-well microplate integrated with a metal-dielectric metasurface (consisting of Teflon substrate and copper microresonators). The well of the microplate was used as a tested liquid holding volume. The microplate is made of polycarbonate and polyamide, chosen for their low dielectric permittivity and minimal dielectric loss such as $\varepsilon'_{pc}$=2.9, $\varepsilon''_{pc}$=0.01 and $\varepsilon'_{pa}$=3.5, $\varepsilon''_{pa}$=0.0027, respectively. The integration of the two structures was modeled as a metasurface based sensor for IgG concentration determination in aqueous solution. The resonant metasurface design was constructed using the finite-element method to solve Maxwell's equations and implemented with usage of COMSOL Multiphysics software, version 6.0 (license number 17078683). In the modeling the metasurface was exposed to TE- and TM- polarized external electromagnetic field with Floquet periodic boundary conditions applied to a unit cell.

## RESULTS AND DISCUSSION

**Complex permittivity of IgG aqueous solutions**

At the first stage of our study we performed experimental probing of IgG aqueous solutions with different concentrations using the microwave dielectrometry method [18]. The experimentally obtained dependencies of the real and imaginary CP values on IgG concentration in water solutions are presented in Fig. 1. The obtained experimental data reveal that the IgG concentration in solutions increase leads to CP values decrease. This result is in a good agreement with the effect which we observed earlier in the microwave dielectrometry study of human serum albumin aqueous solutions [11]. This effect can be explained by the decrease in the amount of free water molecules with the IgG concentration growing in the solutions as a result of the protein molecules hydration process. Similar results were also reported in other previous investigations [19], [20]. The experimentally obtained CP values of the IgG water solutions were used at the next stage of the current study for metal-dielectric metasurface numerical modeling at microwaves.

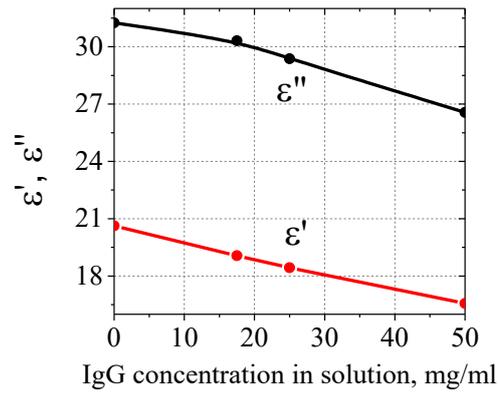

Fig. 1. Dependence of the real and imaginary CP parts on concentration of IgG in aqueous solution at 31.82 GHz at the temperature of 17˚C.

**Numerical modeling of the metal-dielectric metasurface for the IgG determination in aqueous solutions**

The proposed idea of detecting changes in analyte concentration in aqueous solution using the metasurface sensitive structures is based on the effect of a shift in the resonant frequency of the wave reflection spectrum for this metasurface structure contained the tested solutions. In the current study the resonant metal-dielectric metasurface was developed and presented in Fig. 2a. The full wave modeling was conducted for the metasurface depending on the number, position and geometrical sizes of the microresonators in the unit cell. Fig. 1b illustrates the magnitude of the wave transmission spectra for our metal-dielectric metasurface with varying numbers of microresonators. The distribution of the electric and magnetic field components of the metasurface was analyzed. In Fig. 2c demonstrates the colour map of the magnetic and electric field distributions for the first and second resonances (modes) of the metasurface at $f_1$=35.6 GHz and $f_2$=41.27 GHz which consists of one or four microresonators in the unit cell with a liquid (water) layer thickness of $L$ = 0.12 mm for $m$ = 0.5 mm and $q$ = 1.68 mm. In the frequency range of the transmission spectra $S_{12}$ from 33 to 42 GHz, two modes are observed for the metasurface with one microresonator in the unit cell. The same two modes are observed in the transmission spectra $S_{12}$ of the metasurface with four microresonators in the unit cell, as indicated by a similar distribution of the electromagnetic field components around the central microresonator (Fig. 2c). The analysis revealed that the presence of multiple resonators significantly alters the field distribution, enhancing the field intensity. At the frequencies of the first resonance of the transmission spectra $S_{12}$, the electric field is concentrated outside the smaller sides of the central microresonators, while the magnetic field is concentrated in the central region of the larger sides of the microresonators (Fig. 2c). The colour maps of the electric and magnetic field components demonstrate that increasing the number of microresonators from one to four significantly enhances the electromagnetic field intensity. This indicates that the size and number of resonators plays a crucial role in tuning the resonant characteristics and optimization of the metasurface structure. We revealed the metasurface with four micrioresonators in the unit cell as an optimal structure in the following modeling.

A similar metasurface structure was used to investigate the effect of two-mode plasmon-induced transparency in transmission spectra [21]. In the current study, we integrated such a metasurface with the multi-well laboratory microplate structure contained the tested solution to develop the sensor based on the metasurface properties to amplify the electromagnetic waves. In this work, we obtained a similar frequency dependence of the reflection $S_{11}$ and transmission $S_{12}$ spectra of the developed metasurface structure, as well as a similar distribution of the components of the electromagnetic field, which was observed in the earlier study of electromagnetically induced transparency of the metasurface [21].

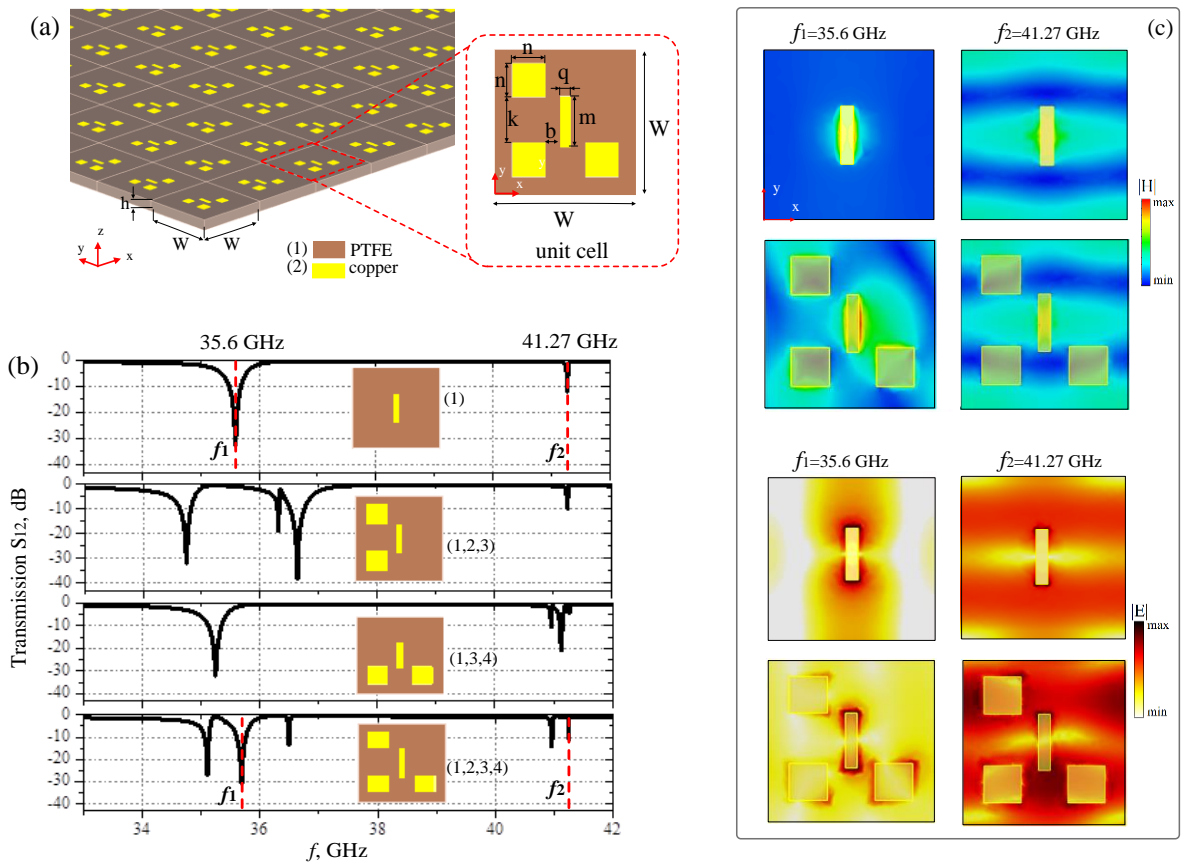

Fig. 2. (a) Perspective view of the metasurface and top view of the metasurface unit cell with four microresonators. The unit cell has a period $W = 7$ mm, with a square size $q = 0.5$ mm, $m = 2.4$ mm, $n = 0.5$ mm, $d = 2$ mm, and thickness of the copper microresonators $t = 0.035$ mm (microresonators are presented as the yellow colour elements in brown unit cell). (b) Magnitude of the wave transmission spectra for the metal-dielectric metasurface with varying numbers, forms and location of microresonators in the unit cell. (c) Colour map of the magnetic and electric field distributions for the first $f_1 = 35.6$ GHz and second $f_2 = 41.27$ GHz resonances (modes) of the metasurface consisting of a single and four microresonators in the unit cell.

The standard all-dielectric 96-wells microplate was integrated with the metal-dielectric metasurface with optimized unit cell geometry to determine the concentration of IgG in aqueous solutions. The structure and S-parameters of this metasurface are illustrated in Fig. 3. The unit cell has a period $W = 8.75$ mm with geometrical sizes of $q$ from 0.35 to 0.5 mm, $m$ from 0.5 to 2.4 mm, $n = 0.5$ mm, $d = 2$ mm, $b = 2.035$, $D = 6.75$ mm, $h = 1.2$ mm, $k = 1.5$ mm, $H = 12$ mm, and with the thickness of the copper plates $t = 0.035$ mm. The parameters stay the same unless specified otherwise.

When the modeling includes the metasurface consisting of the metal-dielectric part integrated with all-dielectric part (the 96-wells microplate) with water, the first ($f_1$) and second ($f_2$) resonance frequencies in the $S_{11}$ reflection (Fig. 3 c) and $S_{12}$ transmission (Fig. 3 d) spectra are observed. However, the resonance frequencies shifted towards lower frequencies ($f_1$ from 35.6 to 26.55 GHz; $f_2$ from 41.27 to 29.88 GHz) due to changes in the unit cell structure including increase of the unit cell period W from 7 to 8.75 mm. This indicates that the integration of these parts of metasurface maintains the modes behaviour similar to observed in Fig. 2b.

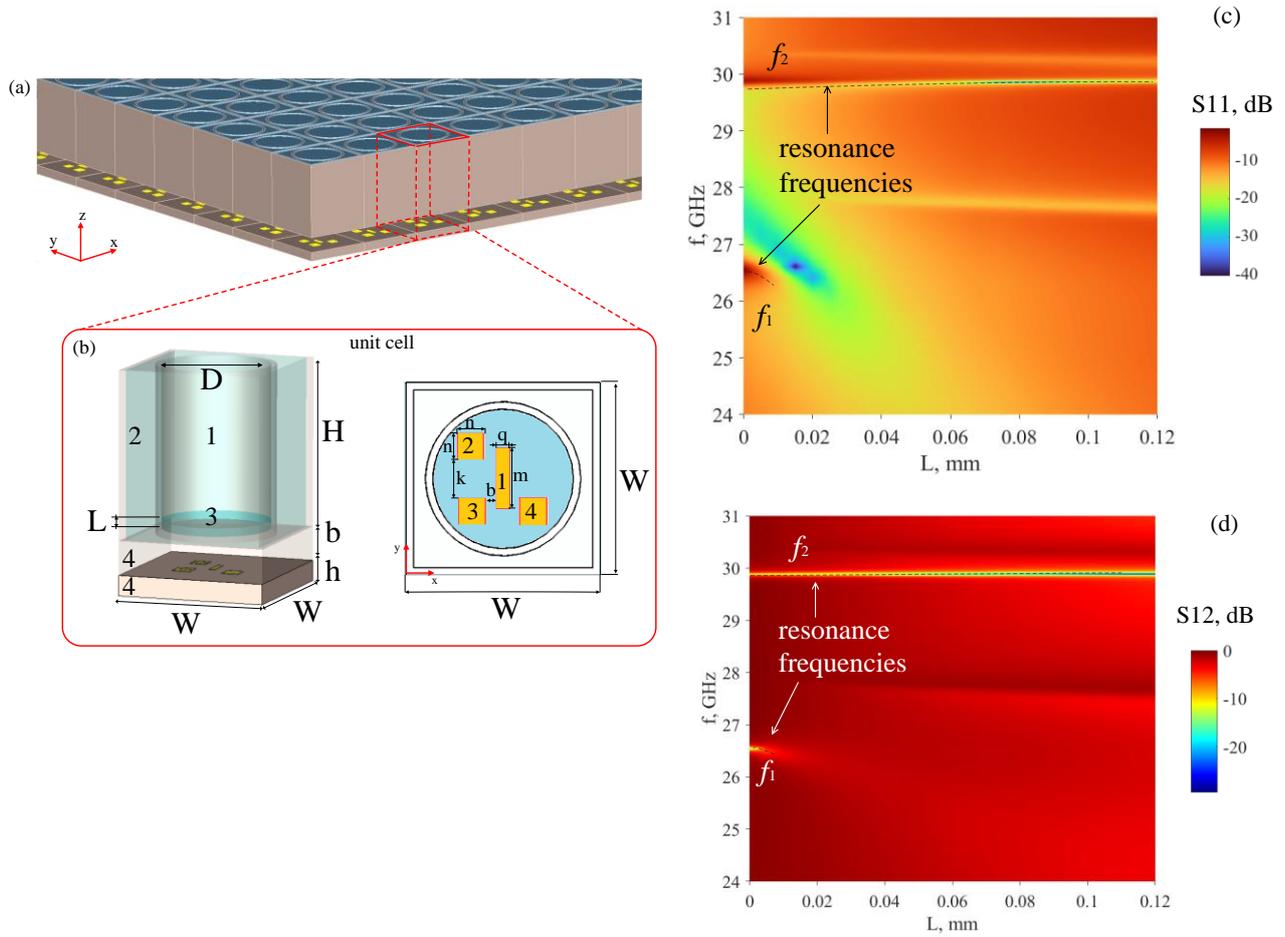

Fig. 3. The metasurface sensory structure based on electromagnetically induced transparency of the metasurface (the unit cell has a height of $h$ containing four microresonators) integrated with the multiwell laboratory microplate: perspective view of the metasurface (a) and the unit cell in the insert (b); 1 – polycarbonate, 2 – polyamide, 3 – the tested liquid layer $L = 0.12$ mm, 4 – Teflon, yellow colour – copper microresonators. Magnitude of the wave reflection $S_{11}$ (c) and transmission $S_{12}$ (d) spectra for the metal-dielectric metasurface with varying numbers, forms and location of microresonators in the unit cell (microresonators are presented as the yellow colour elements in brown unit cell).

The ability of the similar metasurface sensoring structures to determine the human serum albumin concentration in aqueous solutions was validated by numerical modeling in our previous studies, with the maximum sensitivity determination of the albumin concentration as 0.075 MHz/(mg/ml) [20].

To achieve higher sensitivity in determining the IgG concentration in the aqueous protein solutions, in the current work we conducted a parametric study on the influence of the central microresonator size parameters $m$ and $q$ on the Q-factor of the resonance frequency of the reflection coefficient $S_{11}$. The results of the metasurface unit cell optimization are presented in Fig.4. Fig 4 shows the Q-factor (Q-factor=$f_0/\Delta f$, $f_0$ represents the resonance frequency, while $\Delta f$ denotes the full width at half maximum of the resonance) of the resonance frequency of the coefficient $S_{11}$ for different values of size of the central microresonator $m$ and $q$. Parameters of the unit cell that are unchanged: $n = 1$mm, $b = 2.035$ mm, $d = 1$ mm, $L = 0.12$ mm. The study found that the proposed metasurface has the highest sensitivity when the value of $m$ approaches to the value of k (distance between the second and third microresonators). Within the range of $m$ from 1.6 to 1.7, the resonance in the $S_{11}$ reflection spectra was observed with the Q-factor greater than 1500. This indicates that when $m$ is close to $k$, the structure exhibits the high Q-factor of the resonance frequency. Additionally, for $q$ values between 0.35 and 0.5, a similar high-Q resonance was observed in the $S_{11}$ reflection spectra.

Fig. 5a demonstrates the Q-factor of the resonance of the reflection coefficient of the metasurface for different numbers of microresonators in the unit cell with liquid layer thickness of $L = 0.12$ mm. The highest Q-factor of the resonance frequency of the reflection spectra was observed for the metasurface with four micro resonators in the unit cell. The optimized metasurface design is characterized by high sensitivity, making it suitable for exploring the interaction between tested substances and microwaves.

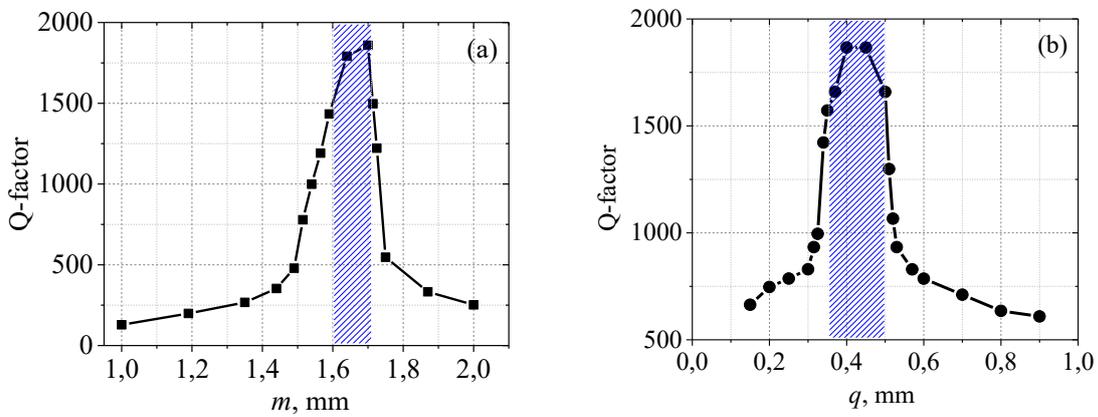

Fig. 4. Q-factor of the resonance frequency of $S_{11}$ reflection coefficient for the metasurface (the unit cell of which contains 4 microresonators with positioning similar to that in Fig. 3b) dependency on the value of the central microresonator size $m$ (a) or $q$ (b) with the fixed other value $q = 0.35$ mm (a) and $m = 1.53$ mm (b).

Fig. 5b illustrates calculated results of the reflection spectra $S_{11}$ of the developed metasurface with different concentration of IgG in the solution. The data presented in Fig. 5b demonstrate that with increase in IgG concentration in water solutions, the calculated reflection spectra minima shift to the lower frequencies. It can be observed that the resonance frequency shift is higher for the unit cell of the metasurface with four microresonators compared to the one microresonator. Therefore, the number of microresonators is a decisive factor in determining the sensitivity of the metasurface for detecting the concentration of IgG in the water solutions. The decrease in the amplitude of the resonance frequency of the reflection coefficient with increasing of the protein concentration in the solution for structure 2 (Fig. 5b) can be explained by decreasing of the imaginary part of the dielectric permittivity (Fig. 1c). The imaginary part of the dielectric permittivity is responsible for the losses in the substance, so as its value decreases due to decrease of the losses in the substance. This leads to a decrease in the amplitude of the resonance frequency, which is observed on the graph of the dependence of the resonance frequency of the reflection coefficient on the frequency for different IgG concentrations Fig. 5b.

Fig. 5c shows the dependences of the resonance frequency shift of the wave reflection spectra ($\Delta f_{S11}$) with increase of the IgG concentration in the solution comparing with pure water. These dependencies can be considered as calibration graphs for the IgG level determination in the aqueous solutions. It can be seen that the value of $\Delta f_{S11}/\Delta C_{IgG} = 0.248$ MHz/(mg/ml) is highest for the metasurface whose unit cell contains 4 microresonators (Fig. 5c). The observed resonance frequency shift in the reflection spectra for various concentrations of IgG in the solutions demonstrates the high sensitivity of the metasurface to changes in concentration of IgG in tested solution.

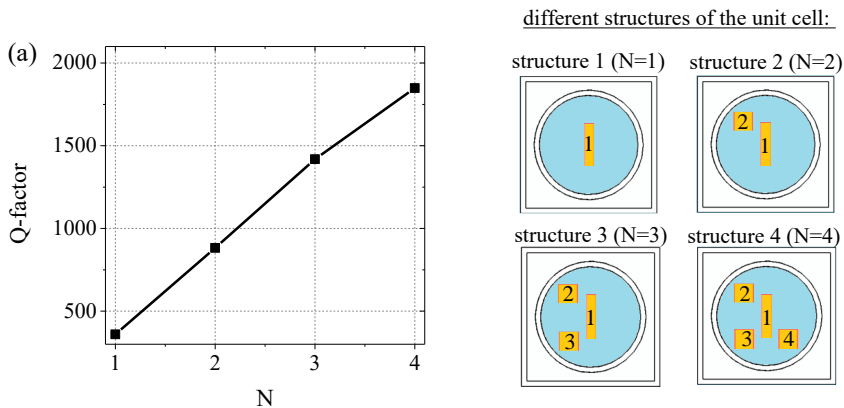

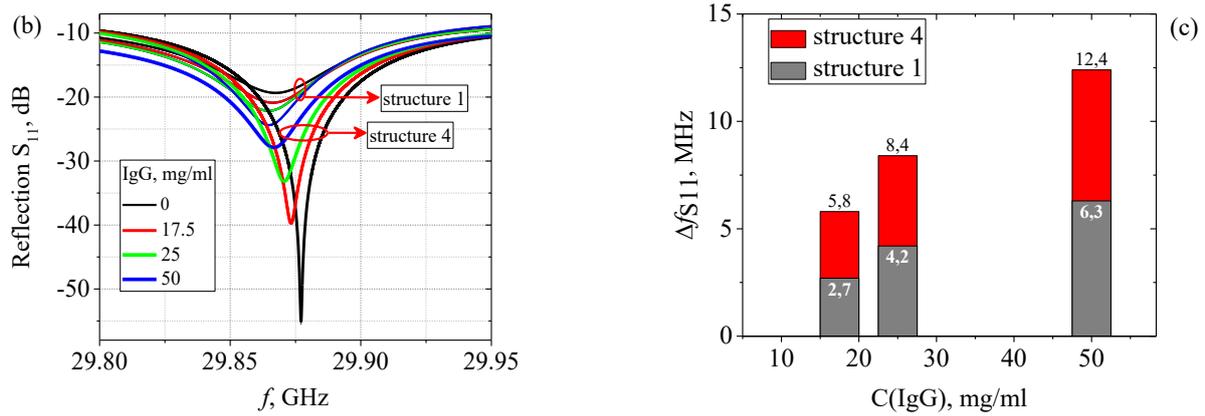

Fig. 5. (a) Q-factor of the resonance of the reflection coefficient of the metal-dielectric metasurface for different numbers N of microresonators in the unit cell - top view; (b) Calculated results of the reflection spectra $S_{11}$ of the metal-dielectric metasurface integrated with the multiwell laboratory microplate with the IgG solutions for different number of the micro resonators in the unit cell ($L = 0.12$ mm); (c) Dependence of the resonance frequency shift of the wave reflection spectra $S_{11}$ of the metal-dielectric metasurface on the IgG concentration in aqueous solutions.

## CONCLUSIONS

The optimized structure of the metal-dielectric metasuface for IgG concentration determination in solutions is proposed based on the results of the experimental probing of the IgG aqueous solutions by differential microwave dielectrometry as well as numerical modeling of the metasurface by finite element method. In the experimental measurements of the IgG solutions of different concentration it was established the dependence of the solutions CP on the protein concentration. Increase of the IgG concentration resulted in the decrease of the CP values of the liquid samples tested due to decrease of the amount of free water molecules in the solutions as a result of the protein molecules hydration growing with the growing of the protein concentration. The experimentally obtained CP values for the IgG aqueous solutions were used in the numerical modeling of the metal-dielectric metasurface unit cell. The standard multi-well microplate was integrated with the metal-dielectric metasurface consists of copper microresonators with Teflon substrate. The optimization of the structure of the metasurface unit cell was conducted with the purpose to enhance the metasurface sensing sensitivity. The modeling results demonstrate that the size and position of the metal microresonators within the unit cell significantly impact sensitivity of the developed metasurface structure for determining the concentration of IgG in aqueous solutions. It was found that the highest Q-factor of the resonance frequency of the reflection spectra was observed for the metasurface with four micro resonators in the unit cell. The optimal parameters of sizes of the central microresonator of the metasurface unit cell $m$ ranging from 1.6 to 1.7 and $q$ from 0.35 to 0.5 were found, providing the highest Q-factor of the resonance of reflection spectra of the metasurface. These optimized parameters enhance the sensing capabilities of the proposed metal-dielectric metasurface, enabling the detection of IgG concentration in aqueous solutions through the resonance frequency shift $\Delta f_{S11}$ in the reflection spectra. The sensitivity value of $\Delta f_{S11}/\Delta C_{IgG}$ is equal to 0.248 MHz/(mg/ml) for the interval of IgG concentrations of $\Delta C_{IgG}$ from 0 to 50 mg/ml. The design of the proposed optimized metasurface based sensing structure may be applied for development of the sensor for non-destructive measurement of IgG and other biological active substances in the liquid samples.

## ACKNOWLEDGMENT

K. S. Kuznetsova acknowledges the support provided by "The Pauli Ukraine Project", funded by the Wolfgang Pauli Institute Thematic Program "Mathematics-Magnetism-Materials" (2023/2024). Z. E. Eremenko acknowledges the funding from the European Union under the Marie Skłodowska-Curie grant agreement no. MSCA4Ukraine project number 1.4 - UKR - 1232611 - MSCA4Ukraine (IFW Dresden). 1.4 - UKR - 1232611 project has received funding through the MSCA4Ukraine project, which is funded by the European Union. Views and opinions expressed are however those of the author(s) only and do not necessarily reflect those of the European Union, the European Research Executive Agency or the MSCA4Ukraine Consortium. Neither the European Union nor the European Research Executive Agency, nor the MSCA4Ukraine Consortium as a whole nor any individual member institutions of the MSCA4Ukraine Consortium can be held responsible for them.